# Why Economic Theories and Policies Fail?
# Unnoticed Variables and Overlooked Economics

Victor Olkhov

Independent, Moscow, Russia

victor.olkhov@gmail.com

ORCID: 0000-0003-0944-5113

**Abstract**

Accuracy of economic theories and efficiency of economic policy strictly depend on the choice of the economic variables and processes mostly liable for description of economic reality. That states the general problem of assessment of any possible economic variables and processes chargeable for economic evolution. We show that economic variables and processes described by current economic theories constitute only a negligible fraction of factors responsible for economic dynamics. We consider numerous unnoted economic variables and overlooked economic processes those determine the states and predictions of the real economics. We regard collective economic variables, collective transactions and expectations, mean risks of economic variables and transactions, collective velocities and flows of economic variables, transactions and expectations as overlooked factors of economic evolution. We introduce market-based probability of the asset price and consider unnoticed influence of market stochasticity on randomness of macroeconomic variables. We introduce economic domain composed by continuous numeric risk grades and outline that the bounds of the economic domain result in unnoticed inherent cyclical motion of collective variables, transactions and expectations those are responsible for observed business cycles. Our treatment of unnoticed and overlooked factors of theoretical economics and policy decisions preserves a wide field of studies for many decades for academic researchers, economic authorities and high-level politicians.

JEL:

This research received no support, specific grant or financial assistance from funding agencies in the public, commercial or nonprofit sectors. We welcome funding our studies.

# 1. Introduction

Aspirations to find "correct" economic theory that should describe and predict markets growth, price change, economic development, employment demand and etc., etc., result endless economic research. The similar aspirations call for "genuine" economic policy that could solve all problems, improve all economic failures and give birth for economic prosperity. Both tied – the economic policy is verified by economic theory and the economic theory is permanently adjusted by economic policy. What is the possible origin of both endless failures?

Early economic studies (Cantillon, 1755; Cournot, 1838; Clark, 1915) look absolutely contemporary and discuss the same issues and problems that are considered by modern researchers. "The Price and Intrinsic Value of a Thing; The Circulation and Exchange of Goods and Merchandise; Production carried on by Entrepreneurs at a risk; Money and Interest; Market Prices; The Circulation of Money; International Trade and Business Cycles; Banks and their Credit" – these are the Chapters of Cantillon (1755) written 265 years ago. "On the Law of Demand; On the Influence of Taxation on Commodities; On the Competition of Producers; On the Communication of Markets" – are the Chapters of Cournot (1838). "Wealth and its Origin; Wages; The Law of Interest; Economic Dynamics" – are the Chapters of "Essentials Of Economic Theory" by Clark (1915). Since then, studies of economic and financial theory move far ahead (Keynes, 1936; Burns, 1954; Blaug, 1985; Dimson and Mussavian, 1999; Vines and Wills, 2018), but the catalog of economic subjects and the list of economic and financial variables under consideration remain almost the same. For sure contribution to the development of the economic and financial theory was made by numerous contemporary researches (Muth, 1961; Sharpe, 1964; Lucas, 1972; Leontief, 1973; Diebold, 1998; Campbell, 2000; Cochrane, 2001; Wickens, 2008; Shubik, 2011; Hansen, 2013). The list of references for sure doesn't include many researchers but serves as illustration of great progress in economic and financial theories during last decades.

The main subject of almost all economic theories concern investigation and modelling factors those impact evolution of few economic and financial variables: asset prices, economic growth, investment, business cycles, inflation and employment, economic risks and uncertainty, demand and supply relations and so on. Most economic policies pretend solve the same economic problems as well as all other, but quickly, chop-chop, before the next election time term (Blinder, 2019). Economic problems are really tough and allow various treatments and numerous solutions. However, above important economic problems don't deplete the questions, cases and issues that should be studied to make economic theory and policy more consistent with economic reality.

We propose that one of the reasons of the long-term failure of economic theories and policies may be explained by the gap between the manifold of properties and processes that conduct real economics and the "short" list of economic variables and relations described by current economic models. Indeed, there are almost no chances to reach the ambition to model, forecast and mange real economics if one considers and uses only 2-3% of existing economic parameters and processes.

We believe that it is reasonable consider general relations and issues that arise in real economics without hasty attempts select "key" economic variables and then derive their impact on economic notions of particular interest. Even politicians know that "key" issue today could be useless tomorrow. Thus excess attention of researchers to "key" economic properties and relations only has not much sense. We propose assess the list of economic variables and properties those describe economic evolution and study their general mutual relations. The number of economic and financial variables under consideration is enormous and many types and classes of economic variables and parameters elude attention of researchers, business and politicians. Such omissions cause wrong economic predictions, false investor's expectations and mistaken economic policy followed by great financial losses of entire economy.

In our paper we consider economic and financial variables and processes on equal basis without selection "key" or "significant" variables. We regard unnoticed economic parameters that are responsible for description of economic dynamics and uncertainty. Both are of great importance and we don't waste time explaining why.

Macroeconomic dynamics is determined by change of macroeconomic variables. It is of great importance to study the economic origin of factors those form and change macroeconomic variables. Econometrics provides perfect methods for assessment of the macroeconomic variables (Fox, et al. 2017). However, considerations of the processes those govern composition of the macroeconomic variables may deliver many surprises. Currently one considers economic variables of separate agents, particular industry and variables that describe macroeconomics as a whole, like investment and credits, supply and demand of the entire economy. However we believe that definitions of economic variables should reflect certain economic approximation. In this paper we propose definitions of economic variables those depend on the choice of numeric scales. It is impossible "exactly" describe real economy. Each economic theory and model should provide certain approximation, simplification of the real economy. We argue why and how economic variables and processes determined by different numeric scales describe approximations of the economic evolution.

Economics is a system with strong internal ties. Dynamics of the economic variables depend on economic uncertainty. Roots of economic uncertainty and ties with uncertainty of financial markets and market trade stochasticity establish complex puzzle that worth investigation. We consider the roots of economic uncertainty and uncover overlooked relations between uncertainty of financial markets and macroeconomic uncertainty. To explain these relations we complement the existing economic variables by additional sets of unnoticed variables and processes those describe uncertainty of financial markets and economic uncertainty as a whole.

In the next section we consider the nature of economic variables. Economic agents perform all economic transactions under action of different risks. We propose replace current letter notations of risk grades by numeric continuous risk grades. We take that risk grades of *n* risks fill unit cube $[0,1]^n$ in $R^n$ and call it economic domain. We introduce economic continuous media approximation in economic domain and describe collective economic variables as functions of risk coordinates. Transition from letter risk grades to numeric continuous risk grades uncovers unnoticed motions of economic agents in economic domain that generate flows of economic variables. In Sec. 3 we consider description of collective market trades and collective agents expectations and their flows in economic domain. We demonstrate that unnoticed flows of collective economic variables and market trades in bounded economic domain $[0,1]^n$ in $R^n$ give origin of oscillations those usually notated as business cycles. In Sec.4 we consider market roots of economic uncertainty and introduce new market-based probability of the market price. Conclusion - in Sec.5. We believe that readers are familiar with modern economic theories, preliminaries of probability theory, partial differential equations and etc. We use roman *A, B* to denote scalars and bold ***P, v, x*** for vector variables.

**2. Economic agents and variables in economic domain**

Any economic theory describes processes and relations those obey approximations determined by definite scales. In particular, any economic theory approximates evolution of variables and relations between them that are averaged or aggregated during selected time interval *Δ*. For example, economic model can approximately describe variables aggregated during interval *Δ* that equals week, month, quarter, etc. Theory with averaging interval *Δ* that equals 1 month doesn't model variations during hour, day or week. However, choice of time averaging interval *Δ* is not sufficient to develop adequate model of real economic processes.

One should consider at least one more scale – "space" scale that determines aggregation of economic agents and their variables by "space" parameter.

Actually, common treatment of the economic variables those describe industry sectors is based on aggregation of variables, like energy consumption, investment, labor etc., of economic agents those belong to particular industry (Fox, et.al., 2017). Variables those describe entire economy are composed by aggregation of all economic agents. Economic parameters also can be formed by aggregation of agent's variables those belong to similar age range, wealth level, gender, geographical position and etc. All these cases consider economic agents as initial bricks for definition of economic variables. Economic agents are distributed by parameters like industries, age, wealth, gender and etc. We would like attract attention to other parameters that are in use for decades. Indeed, during at least four decades major risk rating companies (S&P, 2014; Fitch, 2018; Moody's, 2018) assess credit risk rating of large international banks and distribute banks by their risk grades. Each rating agency utilizes its own risk rating methodology and uses own risk grade system. Risk grades are noted by letters like: *AAA, AA, BB*, and etc. and list of risk grades may have 30-40 different letter notations.

Moreover, rating agencies assess transition probabilities. Rating agencies assess probabilities that during selected time term (6 months, 1 year, etc.) credit rating of the particular bank can change from the current value to a different one and become more secure or more risky (Metz Cantor, 2007; Moody's, 2009; Fitch, 2017; S&P, 2018). Simply speaking, risk transition matrices describe probable motion of economic agents – large international banks – from one risk grade to another during the selected time term.

Above examples confirm that attributing economic agents by parameters like: particular sector of economy, age or wealth range, geographical position or risk rating – is a well-known conventional tool of business and economic research for decades. Attributing agents by parameters allow aggregate economic agents by different ranges of these parameters and develop different approximations of economic evolution.

However, agent's parameters mentioned above have certain common flaw. Indeed, their usage doesn't allow enjoy benefits those can be achieved by description of economic agents and their variables by parameters or coordinates of some metric space that reflects properties of macroeconomics. We believe that usage of geographical positions of agents as "metric space" has almost no economic meaning.

That is the key point of our treatment of theoretical economics. It is habitual that economic agents can be considered as primary bricks of the economic relations. Variety of economic agents and diversity of their economic and financial variables encourage aggregation of agents by some parameters to assess their collective variables. That inspires distribution of agents by industries or economic sectors to assess collective variables of sectors. Such approximation helps model mutual impact of economic variables of different industry sectors. Distribution of agents by industries allows Wassily Leontief develop his Input-Output Analysis as a model of the world economy (Leontief, 1955; 1973). Distribution of agents by parameters as wealth, age, risk ratings and etc., allows study collective variables of groups of agents. Introduction of parameters those help aggregate agents within certain range of parameters establishes ground for definition of collective economic variables of group of agents. For example, collective energy consumption of agents those belong to same industry determines energy consumption of the particular industry.

We believe that usage of risk ratings in a way similar to the current common practice (Fitch, 2017; 2018; Moody's, 2009; 2018; S&P, 2014; 2018) can help describe economic evolution in economic metric spaces. Assessment of risk ratings of large banks – economic agents, can be treated alike to assessment of coordinates of economic agents in metric space. To do that the current methodologies for assessments of risk ratings should be upgraded. Indeed, the key issue of risk assessment is the notation of risk grades. Current notations of risk grades serve the only goal – they protect the business model of the particular rating agency. Risk grade systems of different agencies have different letter notations and agencies use different number of grades. Each agency has its own risk rating methodology. Business of Fitch, Moody's and S&P to a great extent is determined and protected by their specific methodology, risk grade notations and reputation. However, letter notation of risk grades is not the only way denote the value of risk grade and for sure is not the best one. Almost 80 years ago Durand (1941) and later Myers and Forgy (1963) proposed numerical credit grade systems. Actually, there are no any obstacles in transition from letter grade to numerical grade system. There is absolutely no difference how one note risk grades: as *AAA, A, BB* or as *1, 3, 5*. The only impact of transition from letter to numerical grades concerns unification of methodologies of different risk rating agencies. Only prevention of major rating agencies in their ambition to protect their business hinders the transition from letter to numerical risk grade systems. In our opinion the effect of transition will be the opposite. Transition from letter grades to numerical grades system will boost the world risk rating business by much more than 100 times. We assume that the first enterprise that will understand that issue can take over the multi-hundred billions market.

Actually, one can consider letter credit risk grades *AAA, AA, BB*, etc., as points $x_1, x_2,...$ on numerical axis. Risk grades have conditional meaning and one can always take grades as points $x_1$, $x_2$,... of the unit interval *[0,1]*. One can take most secure grade *AAA* to be equal zero and most risky grade to be equal 1. Unification of different risk grade methodologies of major rating agencies allows project their letter risk grades as unified numerical grades of the unit interval [0,1]. Such transition is not the problem of economics or risk management, but only the problem of common agreement between major rating agencies.

We propose introduce numeric **continuous** risk grades those fill interval *[0,1]* instead of discrete numeric grades. Usage of continuous risk grades is also the problem of the methodology of risk assessment only. There are absolutely no economic obstacles that prevent usage of numeric continuous risk grades. We avoid here discuss unified risk methodology. We are sure that high professional teams of researchers of rating agencies or team as Fox, et al. (2017) can solve that problem and develop reasonable rating methodologies for assessment of risk rating of different economic agents under action of different risks. We propose that unified methodology can deliver assessment of ratings for all economic agents of the entire economy for different economic and financial risks those affect economic evolution. We show below, that the proposed transition from letter risk grades to numeric continuous risk grades gives great advantages for the development of the theoretical economics, description of business cycles, economic fluctuations, financial markets, economic forecasting and etc. Usage of continuous risk grades establishes new approach to the description of the economic evolution, financial markets and uncovers unnoticed tough hurdles and overlooked economic notions on the way to "correct" economic theory.

*Economic domain - continuous risk grades*

Economic agents always act under action of economic, financial, market, technological, political etc. risks. Moreover, agents generate risks themselves. Agents economic, market, investment, credit, production, political, technological decisions generate almost all risks. We consider risks as essential part, as integral condition of any economic and financial processes. We believe that economic growth is possible under the action of risks only. The disappearance of risk will initiate halt of all economic processes and destruction of the macroeconomics. Numerous risks impact agents and economic processes. Some of risks have more effect on economic processes than others. It seems almost impossible take into account and estimate risk ratings of all possible risks. The choice of one, two, three most influential risks and assessments of agents' ratings under action of these risks give definite approximation of the economic processes. Continuous risk ratings of the single risk fill the unit interval [0,1] and ratings of two or three most influential risks fill the unit cubes in the $R^2$ or $R^3$ space. Thus, economic agents under the action of 2-3 risks move inside the unit cubes in the $R^2$ or $R^3$ space. Indeed, agents' economic activity, market, technology and numerous other factors impact entire economy and force change of agents' risk ratings. Such "motion" of risk rating of large banks is already described for decades by risk transition matrices provided by rating agencies (Metz Cantor, 2007; Moody's, 2009; Fitch, 2017; S&P, 2018). Usage of continuous risk grades helps clarify agents' risk motion. Let's take that element of risk transition matrix $a_{ij}$ describes probability of transition of particular agent from risk rating $x_i$ to rating $x_j$ during time term *T*. Then one can assess probable velocity $v_{ij}$ of agent's risk motion from $x_i$ to $x_j$ with probability $a_{ij}$ in economic domain during time *T* and mean risk velocity $v_i$ of agent at point $x_i$ as:

$$\boldsymbol{v}_{ij} = \frac{x_j - x_i}{T} \quad ; \quad \boldsymbol{v}_i = \sum_j a_{ij}\, \boldsymbol{v}_{ij} \quad ; \quad \sum_j a_{ij} = 1 \qquad (2.1)$$

We remind, that current notations of letter risk grades don't allow interpret the difference between two grades $x_j$ - $x_i$ as a length (2.1). Hence one can't introduce such a notion as a mean velocity $v_i$ (2.1) of agent with risk rate coordinate $x_i$. Change of notations as proposed by Durand (1941) and Myers and Forgy (1963) opens a new look on description of evolution of the economic agents and the entire economy under action of *n* risks in the economic domain determined as unit cube *[0,1]$^n$*. Dimension *n* of the unit cube defines the number of risks under consideration of particular economic approximation.

And now it is time to remind about *the scales* of the economic approximation. The long road above now led us to important point: for the given number *n* of major risks, economic approximations in the economic domain *[0,1]$^n$* is determined by the choice of two scales: by the time interval *Δ* and by the space scale *l*: 0<*l*≤*1*. The time interval *Δ* determines *scale* of time aggregation and averaging of agents' economic variables and parameters during *Δ* and establishes the divisions of the time axis multiply of *Δ*. The *space scale l* determines aggregation and averaging of economic variables and parameters of agents with risk coordinates near point *x* in a small volume *dV~l$^n$* of the economic domain. Such aggregation makes description of agents by their risk coordinates more roughen by the scale *dV~l$^n$*. Space scale *l* and aggregation by *dV~l$^n$* allows develop economic approximations starting with assumption of imaginable precise assessment of agents' risk ratings and approximate their evolution for *l: 0<l<1* and up to description of the entire economy for *l=1*. Thus *space scales* 0<*l*≤*1* delivers the spectrum, sequence of economic approximations determined by aggregation of agents by different volumes *dV~l$^n$* in the economic domain *[0,1]$^n$*. In a sense, Leontief's input-output analysis is based on division of the entire economy by industries. In a certain manner that division is alike to the division of the economy by volumes *dV~l$^n$* in the economic domain. Both approaches distribute economic agents by different "boxes". Leontief collects agents by "boxes" that define different industries and studies input-output transactions between industries. We collect agents by risk coordinates inside "boxes" $dV \sim l^n$ near risk point *x*, study evolution of their collective economic variables and describe market transactions between agents with coordinates at risk points *x* and *y*. That "small" difference opens wide opportunities for development of the economic theory in the economic domain. The main distinction: we replace aggregation of agents by non-metric "industry boxes" with aggregation of agents inside small "boxes" $dV \sim l^n$ near risk points of metric space $R^n$. That uncovers complex and important unnoticed and overlooked economic variables, relations and processes those define the state and conduct the dynamics of macroeconomics. Disregarding of these hidden variables and processes cause lack of the adequate description of the current state of the economy and incapacity of trustworthy durable economic forecasting.

We consider here main definitions of collective variables and their flows and refer (Olkhov, 2016a-2020) for further details. As we mentioned above, economic decisions of agents, market trade, economic fluctuations and other factors cause the change of agents' risk coordinates. That results in motion of agents with risk velocity $v_i$ (2.1) in economic domain $[0,1]^n$. Motion of particular agent with velocity $v_i$ causes that agent carries his economic variables in the economic domain. Aggregation of agents and their variables near risk point *x* of economic domain inside small "box" $dV \sim l^n$ during time $\Delta$ determines collective variables inside that "box" - volume $dV \sim l^n$. Let $A_i(t_j, x_i)$ denote additive variable *A* of agent *i* with coordinates $x_i$ at moment $t_j$. As additive variable one can consider agent's asset, credit, investment, demand, profits and etc. Sum of additive variables of group of agents equals collective variable of the group. Let take that time series $t_j$ obey (2.2)

$$\Delta = \left[t - \frac{\Delta}{2}; t + \frac{\Delta}{2}\right] \quad ; \quad t - \frac{\Delta}{2} \leq t_j \leq t + \frac{\Delta}{2} \quad ; \quad j = 1, \ldots N \qquad (2.2)$$

Then collective variable *A* of agents inside "box" $dV \sim l^n$ near point *x* averaged during the interval $\Delta$ (2.2) equals

$$A(t, \pmb{x}) = \frac{1}{N} \sum_{j=1}^{N} \sum_{i \in dV(x)} A_i(t_j, \pmb{x_i}) \qquad (2.3)$$

First sum in (2.3) collects variable *A* of all agents *i* with coordinates $x_i$ inside *dV* and the second sum average that value during the time interval $\Delta$ (2.2). Relation (2.3) introduces mean additive variable *A(t,x)* of agents inside *dV* near point *x*. Function *A(t,x)* defines distribution of variable *A* over the economic domain. In some sense (2.3) is similar to distribution of variable *A(t,x)* by different industries, different age, wealth and etc. However, usage of continuous risk grades in economic domain uncovers unnoticed motion of collective economic variables that are overlooked by current economic models. Indeed, as we outlined above, each agent $i$ at point $(t_j,x_i)$ with variable $A_i(t_j,x_i)$ carries that variable with risk velocity $v_i$ (2.1). Collective result of such risk motion for all agents inside "box" $dV$ averaged during $\Delta$ (2.2) defines the collective flow $\boldsymbol{P}_A(t,x)$ that carries collective variable $A(t,x)$ with collective velocity $\boldsymbol{v}_A(t,x)$ (2.4) in the economic domain $[0,1]^n$

$$\boldsymbol{P}_A(t,x) = \frac{1}{N}\sum_{j=1}^{N}\sum_{i\in dV(x)} A_i(t_j,x_i)\,\boldsymbol{v}_i(t_j,x_i) = A(t,x)\boldsymbol{v}_A(t,x) \qquad (2.4)$$

It is important to underline that notions of collective flow and collective velocity have sense even if one considers aggregation of all agents of the entire economy and "box" $dV$ coincides with entire economic domain $[0,1]^n$, so that $dV = [0,1]^n$. In that case macroeconomic variable $A(t)$ takes form

$$A(t) = \frac{1}{N}\sum_{j=1}^{N}\sum_{i} A_i(t_j,x_i) \qquad (2.5)$$

$$\boldsymbol{P}_A(t) = \frac{1}{N}\sum_{j=1}^{N}\sum_{i} A_i(t_j,x_i)\,\boldsymbol{v}_i(t_j,x_i) = A(t)\boldsymbol{v}_A(t) \qquad (2.6)$$

We call the transition from description of economic variables assigned with particular economic agent $A_i(t_j,x_i)$ to description of economic variables $A(t,x)$ (2.3) as functions of coordinates $x$ in economic domain as *economic continuous media approximation*. Flows $\boldsymbol{P}_A(t,x)$ and velocities $\boldsymbol{v}_A(t,x)$ (2.4) of each particular additive economic variable $A$ describe its motion in the economic domain. We consider continuous media approximations for different time and space scales $(\Delta,l)$ as set of intermediate approximations between description of economics as separate agents and macroeconomic approximation derived by collecting variables of all agents of the economy. As we show (Olkhov, 2017c; 2017d; 2018; 2019a-2020), to derive correct description of macroeconomic approximation one should take into account collective flows and velocities (2.6). However, these macroeconomic flows and velocities (2.6) related with numerous different macroeconomic variables $A(t)$ (2.5) are unnoted and overlooked by modern economic theories and that gap definitely causes failures of macroeconomic forecasting.

Relations (2.5) introduce one more important and still unnoticed economic factor – macroeconomic mean risk $X_A(t)$ (2.7) related with particular collective economic variable $A$:

$$A(t)X_A(t) = \frac{1}{N}\sum_{j=1}^{N}\sum_{i} A_i(t_j,x_i)\,x_i \qquad (2.7)$$

It is obvious that motion of macroeconomic variable $A(t)$ with velocity $\boldsymbol{v}_A(t)$ (2.6) as well as motion of mean risk $X_A(t)$ (2.7) in the unit cube $[0,1]^n$ – economic domain can't go beyond its borders. Hence such a motion should follow complex oscillations inside the economic domain. These oscillations of mean risk $X_A(t)$ (2.7) of economic variable $A(t)$, fluctuations of velocity $v_A(t)$ (2.6) as well as fluctuations of collective velocity $v_A(t,x)$ (2.4) inside economic domain reflect processes that are currently noted as business cycles (Olkhov, 2017c; 2017d; 2019a; 2019c; 2020). Different economic variables $A,B,C$ define different collective velocities $v_A(t,x)$, $v_B(t,x)$, $v_C(t,x)$ and different mean risks $X_A(t)$, $X_B(t)$, $X_C(t)$. Their mutual interactions and their oscillations inside the unit cube $[0,1]^n$ of the economic domain establish complex picture of collective fluctuations of macroeconomic variables those observed and treated as business cycles. We consider these overlooked hidden collective oscillations of economic variables, their mean risks and velocities as the origin of observed economic business cycles.

Economic continuous media approximation in the economic domain $[0,1]^n$ uncovers important and unnoticed wave generation and propagation of small disturbances of collective economic variables and market transactions. In economics the term "wave" is in use at least since Kondratieff's waves (Kondratieff, 1935). However, Kondratieff's waves as well as economic cycles describe oscillations of economic variables in time only. To observe and describe economic waves one should consider economic processes in certain space. Introduction of the economic domain as a unit cube $[0,1]^n$ in $R^n$ uncovers possible wave propagation of small disturbances of collective economic variables and market transactions as functions of risk coordinates in the economic domain. We described (Olkhov, 2016a-2017b; 2019c) possible propagation of small waves through the economic domain and a different type of "surface-like" waves those propagate along the borders of the economic domain. We show that possible exponential amplification of the wave amplitudes during the propagation can cause rise of perturbations' amplitudes and result in development of crises processes. Currently, the influence of the economic wave propagation on macroeconomic evolution is completely unnoted.

However, the hidden complexity of the collective economic variables and their flows in economic domain delivers only small fraction of the difficulties on the way for development of the comprehensive economic theory. In the next section we consider unnoticed variables and overlooked economic properties related with the main drivers of the economic development – market trade and expectations.

## 3. Market trade and expectations

Market trade is the origin and generator of the economic development. Markets redistribute the volumes of the existing assets and commodities as well as assets and commodities generated by industrial production over agents in the economic domain. Markets establish the prices of assets and commodities that are adopted by the economy. Each agent involved into particular market transaction takes own decision on the amount of the trade value and volume under personal expectations. Relations between expectations of agents and performance of market transactions establish a complex puzzle for the theoretical economics. Expectations determine preferences and decisions of agents those result in market trade performance. Impacts of expectations on world markets and economic evolution are under research for decades (Muth, 1961; Lucas, 1972; Hansen and Sargent 1979; Blume and Easley 1984; Brock and Hommes, 1998; Manski 2004; Brunnermeier and Parker 2005; Manski, 2017; Farmer, 2019). Expectations generate market trade stochasticity and their measuring and modelling remain the major puzzle of financial economics. Ties between expectations and economic policy (Sargent and Wallace, 1976) result complex impact on world markets and economic development. Methods and description of collective economic variables as functions of risk coordinates in economic domain delivers the unified approach to description of the world markets, economic expectations and economic policy. A first approximation of the relations between market trades and economic expectations described as functions of risk coordinates in the economic domain was developed by Olkhov (2019b; 2019c; 2020). For simplicity in this paper we consider only preliminary, general notions and definitions required for description of the market trade and expectations in the economic domain.

The idea for description of market trade in economic domain is simple and in some sense is alike to Leontief's model (Leontief, 1955; 1973). Leontief collects agents by industries, but we collect agents inside small volume $dV \sim l^n$ near points of economic domain $[0,1]^n$. Leontief considers input-output trade relations between industries. We describe collective market trade between agents those belong to small volumes $dV \sim l^n$ near points $x$ and $y$ of the economic domain $[0,1]^n$. That "small" difference from Leontief's model opens wide opportunities for description collective market trade in the economic domain $[0,1]^n$ of metric space $R^n$. Actually, we reproduce economic continuous media approximation introduced for description of collective economic variables in Sec.2 to model collective market trade and their flows in the economic domain of double dimension. Motion of agents in the economic domain generates flows of collective market trades. Assumptions on relations between market trade and agents expectations allow introduce notions of collective expectations and their collective flows in the economic domain. Notions of collective market trade and expectations as well as notions of their collective flows in the economic domain are unnoticed and overlooked by current economic theories.

Let us explain briefly above argumentation in more details. Let us consider seller and buyer at points $x_i$ and $y_j$ of the economic domain $[0,1]^n$. These two agents at time $t_m$ perform market transaction with a particular asset or commodity at volume $U(t_m,x_i,y_j)$ and value $C(t_m,x_i,y_j)$. Let denote $C(t,x,y)$ and $U(t,x,y)$ as collective trade value and volume between sellers and buyers inside small volumes $dV \sim l^n$ near point $x$ and $y$ averaged during time interval $\Delta$.

$$U(t,\mathbf{z}) = \frac{1}{N}\sum_{m=1}^{N}\sum_{i\in dV(x); j\in dV(y)} U(t_m, \mathbf{z}_{ij}) \quad ; \quad \mathbf{z}=(\mathbf{x},\mathbf{y}) \; ; \; \mathbf{z}_{ij}=(\mathbf{x}_i, \mathbf{y}_j) \quad (3.1)$$

$$C(t,\mathbf{z}) = \frac{1}{N}\sum_{m=1}^{N}\sum_{i\in dV(x); j\in dV(y)} C(t_m, \mathbf{z}_{ij}) \quad (3.2)$$

Transactions between agents at points $x$ and $y$ can be treaded as functions at point $z=(x,y)$ (3.1) and that easily transfers description of the market transactions in the economic domain of double dimension $2n$. Motion of economic agents in economic domain induces corresponding motion of market transactions (3.1; 3.4) in the economic domain of double dimension $2n$. For example, collective flow $P_U(t,z)$ of trade volume depends on collective flows $P_{xU}(t,z)$ of sellers at point $x$ and collective flows $P_{yU}(t,z)$ of buyers at point $y$:

$$\mathbf{P}_U(t,\mathbf{z}) = \left(\mathbf{P}_{xU}(t,\mathbf{z}); \mathbf{P}_{yU}(t,\mathbf{z})\right) \quad (3.3)$$

$$\mathbf{P}_{xU}(t,\mathbf{z}) = U(t,\mathbf{z})\mathbf{v}_{xU}(t,\mathbf{z}) = \frac{1}{N}\sum_{m=1}^{N}\sum_{i\in dV(x); j\in dV(y)} U_{ij}(t_m, \mathbf{z}_{ij})\mathbf{v}_{ix}(t_m,\mathbf{x}) \quad (3.4)$$

$$\mathbf{P}_{yU}(t,\mathbf{z}) = U(t,\mathbf{z})\mathbf{v}_{yU}(t,\mathbf{z}) = \frac{1}{N}\sum_{m=1}^{N}\sum_{i\in dV(x); j\in dV(y)} U_{ij}(t_m, \mathbf{z}_{ij})\mathbf{v}_{iy}(t_m,\mathbf{y}) \quad (3.5)$$

Velocities (3.6) $v_{xU}$ and $v_{yU}$ determined in (3.4; 3.5) as average velocities of sellers and buyers in small volume $dV \sim l^n$ during time interval $\Delta$ with respect to the average trade volume $U(t,z)$ (3.1).

$$\mathbf{v}_U(t,\mathbf{z}) = \left(\mathbf{v}_{xU}(t,\mathbf{z}) ; \mathbf{v}_{yU}(t,\mathbf{z})\right) \quad (3.6)$$

Equations that describe evolution of the collective trade volume $U(t,z)$ (3.1) and value $C(t,z)$ (3.2) depend on flows of the collective trade volume $P_U(t,z)$ and value $P_C(t,z)$. Derivation of these equations and other math methods required for description of the market trade evolution in the economic domain of double dimension are presented in (Olkhov, 2017b; 2017c; 2017d; 2018; 2019b; 2019c; 2020).

*Expectations*

Expectations of agents remain the most mysteries factors of the economic theory. Variety, variability and uncertainty of expectations of agents make them the headache for econometric observations and theoretical economics. Expectations of agents, as "dimension" component of economic theory, are generated by at least other three "dimensions" of economic theory – economic variables, market transactions and economic policy (Olkhov, 2022d). The role of expectations can be treated as a "glue" that ties up impact of the variables, transactions and economic policy and produces market trade decisions of agents.

Even simplified model of mutual dependence of expectations and market trades uncovers unnoticed properties of collective expectations determined by collective trade value and volume (Olkhov, 2019b; 2019c; 2021d). Indeed, according to economic continuous media approximation one should define notions of collective expectations as functions of risk coordinate in economic domain. To do that one should collect expectations of agents in a small volume *dV* and average them during the interval *Δ* (2.2). Sum of different agents' expectations is a tough problem that requires methodological introduction of unified measure of different expectations. That is one of the overlooked problems of current economic studies of agents' expectations and the problem is far from solution. However, if one assumes that the unified measure of expectations is selected and expectations of different agents can be observed and assessed by the unified measure, definition of collective agents' expectations still remains a tough problem. Indeed, different agents' expectations are "responsible" for trade decisions of different value. It seems not fair sum on equal basis expectation 1 that is responsible for trade worth $1 and expectation 2 responsible for trade worth $10MM. We propose that to collect expectations of different agents one should multiply the measure of the expectation responsible for the value of particular trade by the value of that trade. Thereafter, one should multiply by the trade volume the measure of expectation that is responsible for the particular trade volume. These simple rules result definitions of collective expectations as weighted by the market trade value and volume they approved. For example, if $U(t_m, x_i, y_j; k, l)$ at time $t_m$ denotes trade volume between seller at $x_i$ and buyer at $y_j$ (3.1) under sellers expectations $ex_{sU}(t_m, x_i; k)$ of type *k* and buyers expectation $ex_{bU}(t_m, y_j; l)$ of type *l* then collective sellers expectations $Ex_{sU}(t, z)$ of the trade volume take form:

$$U(t,z)Ex_{sU}(t,z) = \frac{1}{N}\sum_{k,l}\sum_{m=1}^{N}\sum_{i\in dV(x); j\in dV(y)} U(t_m, z_{ij}; k, l)\ ex_{sU}(t_m, x_i; k) \qquad (3.7)$$

To avoid excess complexity we omit definitions and equations those describe evolution of collective market trade, collective expectations, collective flows of market trade, collective flows of expectations and their mutual interactions and refer (Olkhov, 2017 - 2020; 2021d).

Withal one should remember that each trade is performed by at least two agents – by seller and by buyer. Expectations of seller and buyer can be different and thus definition of collective expectations should identify collective expectations of sellers and buyers. Thus description of collective expectations those impact market trade with selected asset or commodity requires introduction of four different collective expectations: two expectations those impact the trade value and volume of sellers and two similar expectations of buyers. As we mentioned above, motion of agents in economic domain $[0,1]^n$ cause motion of trades in economic domain of double dimension and that result motion of collective expectations. Flows of collective expectation in bounded economic domain result slow fluctuations that are alike to business cycles of expectations. Econometric observations of such cycles of expectations are absent. However, overlooked impact of expectation's cycles on cycles of market trade and conventional business cycles of economic growth, investment and etc. hide important tools and relations of their management.

We believe that the nature of economic processes, flows of economic variables, market transactions and expectations in the bounded economic domain make absolutely impossible maintaining permanent economic growth without crises and recessions. Nevertheless, econometric observations and theoretical predictions of mutual dependence between collective expectations, market trade and variables could deliver financial authorities, Central Banks and politicians more tools, reasons and argumentations for achieving economic prosperity with less losses.

At the end of this subsection we outline that expectations of agents to a great extend are determined by existing economic, financial, market and etc., laws and regulations. Regularly adjusted political decisions those disturb current economic laws are usually noted as economic and financial policy. Economic laws that conduct market trade, taxes, production and mining, environment protection and etc., – all that amount of economic laws and regulations are permanently adjusted by economic policy. That result unpredictable disturbances of agents expectations those projected into market fluctuations and macroeconomic uncertainty. Modelling of that jurisdictive "black box" as fourth dimension of the economic theory and description of its impact on economic evolution, risk variations, business cycles and etc., - the though problem for future. However, it could be important keep in mind existing of the four dimensions of the economic theory – collective variables, market transactions, expectations and economic policy&laws (Olkhov, 2022d).

## 4. Market roots of economic uncertainty

Randomness of the market trade is studied for decades. In that section we demonstrate that market stochasticity should be treated as important source of economic uncertainty that impact on much more macroeconomic variables and properties than it is recognized now.

Economic and financial transactions between agents determine the major tool that supports economic development and growth. Market trade mysteries broadly define the enigma of the economic theory and forecasting. Agents variables collected by entire economy or by different "boxes" as industries, wealth, age range, risk coordinates and etc., determine the economic state, the starting conditions for the economic development and growth. However, agents variables itself even collected in any manner over the economy unable generate economic evolution. Only market trade establishes and causes the economic and financial development. Only market trade moves the current state of the economy to the future prosperity. One should remember that all theories and forecasts those utilize relations between economic variables only are eluding and fudging description of the real economic ties and laws those conduct economic dynamics. Usual common statements like "supply depends on demand" or "investment depend on bank rate" describe relations between variables, but omit several intermediate and complex chains of market transactions and neglect economic relations those conduct market trade. That is understandable desire to derive predictions under simplified assumptions.

Below we show that such shortening economic relations result in disregarding numerous significant economic properties and variables those conduct market trade and economic forecasts. To start with let us consider as example the market trade of particular asset or commodity. Let denote as $C(t_i)$ and $U(t_i)$ the value and volume of the market trade at time $t_i$ at a price $p(t_i)$, so trivial relations between the trade value, volume and price take form:

$$C(t_i) = p(t_i)U(t_i) \qquad (4.1)$$

Simple trade relations define price $p(t_i)$ of particular trade at $t_i$. Modern market trade is a high-frequency process with high variations of the trade value, volume and price. Most economic and financial models consider averaging and smoothing procedures of market trade time-series during selected time averaging interval $\Delta$. Financial market models can take the averaging interval $\Delta$ to be equal minutes, hours or days and macroeconomic models deal with the averaging interval $\Delta$ to be equal weeks, months and years. Time averaging during the interval $\Delta$ establishes time axis division multiply of $\Delta$ of the theoretical description of economics and financial markets. Different choice of the averaging interval $\Delta$ determines different approximations of the market dynamics and economic evolution.

Let us select the averaging interval $\Delta$ and study statistical properties of the market trade value, volume and price treated as random variables during $\Delta$. Below we explain why and how statistical properties of the trade value and volume determine statistical properties of the market price and how all that define statistical properties of the economic variables.

For simplicity let us take initial market trade time-series at time $t_i$ as multiple of $\varepsilon$ and assume that there are $N$ members of time-series in each interval $\Delta$:

$$t_i = \varepsilon \cdot i \ ; \ \varepsilon \ll \Delta = 2n \cdot \varepsilon \ ; \ N = 2n + 1 \ ; \ i = 0, 1, 2, \dots \quad (4.2)$$

Averaging of time-series $t_i$ during $\Delta$ replaces initial time axis division multiply of $\varepsilon$ by division multiply of $\Delta$ and generates time-series $t_k$ (4.3)

$$t_k = \Delta \cdot k \ ; \ \Delta_k = \left[ t_k - \frac{\Delta}{2}; t_k + \frac{\Delta}{2} \right] \ ; \ k = 0, 1, 2, \dots \quad (4.3)$$

Time-series of the market trade value $C(t_i)$, volume $U(t_i)$ and price $p(t_i)$ during each averaging interval $\Delta_k$ are very irregular. Aggregation or averaging of irregular variables during interval $\Delta_k$ helps derive smooth variables and develop reliable predictions. For simplicity we consider initial time-series of the market trade value $C(t_i)$, volume $U(t_i)$ and price $p(t_i)$ as random variables during each averaging interval $\Delta_k$. Asset pricing theory is the key problem of financial economics and the literature is endless. We refer only few studies (Bachelier, 1900; Kendall and Hill, 1953; Muth, 1961; Sharpe, 1964; Fama, 1965; Karpoff, 1987; Cochrane and Hansen, 1992; Cochrane, 2001; Hansen, 2013). Any reader can add to that list hundreds of his preferred references. Above references confirm only one issue – any new treatment of the asset pricing is very important for economic and financial modelling.

We present derivation of the statistical properties of the market random price time-series during averaging interval $\Delta$ due to Olkhov (2021a-2022d) and refer there for details. Our consideration of the market price probability follows simple thesis: relations (4.1) state that it is impossible independently define probabilities of the trade value $C(t_i)$, volume $U(t_i)$ and price $p(t_i)$ during $\Delta$. Given random properties of the market trade value $C(t_i)$ and volume $U(t_i)$ determine random properties of the price time-series $p(t_i)$ during $\Delta$. In some extend definition of the mean price $p(t_k;1)$ during the interval $\Delta_k$ (4.3) as function of the mean trade value $C(t_k;1)$ and volume $U(t_k;1)$ was given more then 30 years ago as volume weighted average price (VWAP) (Berkowitz et.al, 1988; Buryak and Guo, 2014; Busseti and Boyd, 2015; CME Group, 2020). Definition of VWAP $p(t_k;1)$ (4.4) averaged during interval $\Delta_k$ is based on total value $C_\Sigma(t_k;1)$ and volume $U_\Sigma(t_k;1)$ aggregated during $\Delta_k$ (4.3) and follows (4.1):

$$p(t_k; 1) = E[p(t_i)] = \frac{1}{\sum_{i=1}^{N} U(t_i)} \sum_{i=1}^{N} p(t_i) U(t_i) = \frac{C_\Sigma(t_k; 1)}{U_\Sigma(t_k; 1)} \quad (4.4)$$

$$C_\Sigma(t_k; 1) = \sum_{i=1}^{N} C(t_i) = \sum_{i=1}^{N} p(t_i) U(t_i) \ ; \ U_\Sigma(t_k; 1) = \sum_{i=1}^{N} U(t_i) \quad (4.5)$$

Relations (4.5) define total trade value $C_\Sigma(t_k;1)$ and volume $U_\Sigma(t_k;1)$ during $\Delta_k$ (4.3). Let us note mathematical expectation during $\Delta_k$ as $E[\dots]$. As we assume there are $N$ trades during $\Delta_k$. Hence, mean trade value $C(t_k;1)$ and mean volume $U(t_k;1)$ during $\Delta_k$ equal:

$$C(t_k; 1) = E[C(t_i)] = \frac{1}{N} C_\Sigma(t_k; 1) = \frac{1}{N} \sum_{i=1}^{N} C(t_i) \quad (4.6)$$

$$U(t_k; 1) = E[U(t_i)] = \frac{1}{N} U_\Sigma(t_k; 1) = \frac{1}{N} \sum_{i=1}^{N} U(t_i) \quad (4.7)$$

Thus VWAP $p(t_k;1)$ (4.1; 4.4) averaged during interval $\Delta_k$ (4.3) equals:

$$p(t_k;1) = \frac{1}{\frac{1}{N}\sum_{i=1}^{N} U(t_i)} \cdot \frac{1}{N}\sum_{i=1}^{N} p(t_i) U(t_i) = \frac{1}{\frac{1}{N}\sum_{i=1}^{N} U(t_i)} \cdot \frac{1}{N}\sum_{i=1}^{N} C(t_i) = \frac{C(t_k;1)}{U(t_k;1)} \quad (4.8)$$

These trivial relations (4.4-4.8) uncover important consequence that is usually overlooked. Indeed, relations (4.1; 4.4-4.8) cause zero correlations $corr\{p(t_i)U(t_i)\}=0$ between time-series of the market trade volume $U(t_i)$ and price $p(t_i)$ during averaging interval $\Delta_k$:

$$E[C(t_i)] = E[p(t_i)U(t_i)] = \frac{1}{N}\sum_{i=1}^{N} p(t_i) U(t_i) \equiv \frac{1}{\sum_{i=1}^{N} U(t_i)} \sum_{i=1}^{N} p(t_i) U(t_i) \frac{1}{N}\sum_{i=1}^{N} U(t_i) = E[p(t_i)]E[U(t_i)] \quad (4.9)$$

Hence, from (4.9) obtain (4.10) that $corr\{p(t_i)U(t_i)\}$ equals zero:

$$corr\{p(t_i)U(t_i)\} = E[p(t_i)U(t_i)] - E[p(t_i)]E[U(t_i)] = 0 \quad (4.10)$$

We underline that unnoticed zero correlations between price and trade volume (4.10) is result of the definition of the VWAP $p(t_k;1)$ (4.4 - 4.8). That is an important issue, as studies of the price-volume correlations have long history (Tauchen and Pitts, 1983; Gallant et.al., 1992; Campbell et.al., 1993; Odean, 1998). The choice of the VWAP $p(t_k;1)$ (4.4-4.8) states zero correlations between price and trade volume. Hence studies of the market trade price-volume time-series should be reconsidered with respect to definite price-volume averaging procedures. Economic meaning of price-volume correlations should be considered with respect to economic meaning of the trade volume and price probabilities.

We underline that definition of the VWAP $p(t_k;1)$ (4.4-4.8) is not sufficient to define all random properties of the price $p(t_i)$ during $\Delta_k$ (4.3). To do that one should define all price *n-th* statistical moments $p(t_k;n)=E[p^n(t_i)]$. We introduce all price *n-th* statistical moments as extension of (4.1; 4.4-4.8). For all $n=1,2,3,...$ we take usual *n-th* statistical moments of the trade value $C(t_k;n)$ and total sums of *n-th* power of trade value $C_\Sigma(t_k;n)$ (4.11) and *n-th* statistical moments of the trade volume $U(t_k;n)$ and total sums of *n-th* power of volume $U_\Sigma(t_k;n)$ (4.12) during $\Delta_k$ (4.3) as:

$$C(t_k;n) = E[C^n(t_i)] = \frac{1}{N} C_\Sigma(t_k;n) \quad ; \quad C_\Sigma(t_k;n) = \sum_{i=1}^{N} C^n(t_i) \quad (4.11)$$

$$U(t_k;n) = E[U^n(t_i)] = \frac{1}{N} U_\Sigma(t_k;n) \quad ; \quad U_\Sigma(t_k;n) = \sum_{i=1}^{N} U^n(t_i) \quad (4.12)$$

Taking *n-th* power of (4.1) we obtain:

$$C^n(t_i) = p^n(t_i)U^n(t_i) \quad (4.13)$$

We extend definition of VWAP (4.4-4.8) and for all $n=1,2,3,...$ introduce price *n-th* statistical moments $p(t_k;n)=E[p^n(t_i)]$ (4.14) as:

$$p(t_k;n) = E[p^n(t_i)] = \frac{1}{\sum_{i=1}^{N} U^n(t_i)} \sum_{i=1}^{N} p^n(t_i) U^n(t_i) = \frac{C_\Sigma(t_k;n)}{U_\Sigma(t_k;n)} = \frac{C(t_k;n)}{U(t_k;n)} \quad (4.14)$$

It is obvious, that (4.14) results:

$$C(t_k;n) = p(t_k;n)U(t_k;n) \quad ; \quad C_\Sigma(t_k;n) = p(t_k;n)U_\Sigma(t_k;n) \quad (4.15)$$

To explain and justify our definition (4.14; 4.15) of the market price *n-th* statistical moments $p(t_k;n)=E[p^n(t_i)]$ we mention that we extend VWAP logic for (4.13) as *n-th* power of (4.1). Average *n-th* power of price $E[p^n(t_i)]$ is determined as *n-th* power volume $U^n(t_i)$ weighted average of *n-th* power of price $p^n(t_i)$ (4.14). Both relations (4.14; 4.15) retain the meaning of market *n-th* power of price $p^n(t_i)$ as coefficient between *n-th* power of trade value $C^n(t_i)$ and trade volume $U^n(t_i)$ (4.13). Definitions (4.14; 4.15) maintain meaning of price *n-th* statistical moment $p(t;n)= E[p^n(t_i)]$ as ratio of total sum of *n-th* power of value $C^n(t_i)$ to sum of *n-th* power volume $U^n(t_i)$ during averaging interval $\Delta_k$ (4.3).

One can easy obtain that (4.11- 4.15) cause zero correlations between time-series of *n-th* power of price $p^n(t_i)$ and trade volume $U^n(t_i)$:

$$E[p^n(t_i)U^n(t_i)] = \frac{1}{N}\sum_{i=1}^{N} p^n(t_i) U^n(t_i) = \frac{\sum_{i=1}^{N} p^n(t_i)U^n(t_i)}{\sum_{i=1}^{N} U^n(t_i)} \frac{\sum_{i=1}^{N} U^n(t_i)}{N} = E[p^n(t_i)]E[U^n(t_i)]$$

Hence correlations $corr\{p^n(t_i)U^n(t_i)\}$ between time series of *n-th* power of price $p^n(t_i)$ and trade volume $U^n(t_i)$ equal zero:

$$corr\{p^n(t_i)U^n(t_i)\} = E[p^n(t_i)U^n(t_i)] - E[p^n(t_i)]E[U^n(t_i)] = 0 \qquad (4.16)$$

Zero correlations (4.16) for all *n=1,2,3,…* don't imply statistical independence between price and volume time-series. For example, from (4.14; 4.15) one can easy obtain correlations $corr(pU^2)$ between price $p(t_i)$ and squares of trade volume $U^n(t_i)$ (Olkhov, 2022c). For brevity we omit $t_i$ and obtain:

$$E[pU^2] = E[p]E[U^2] + corr\{pU^2\} \quad ; \quad E[pU^2] = E[CU] = E[C]E[U] + corr\{CU\}$$
$$corr\{pU^2\} = corr\{CU\} - p(t_k;1)\sigma^2(U) \qquad (4.17)$$

Here we denote market trade volume volatility during $\Delta_k$ (4.3) as $\sigma^2(U)$:

$$\sigma^2(U) = U(t_k;2) - U^2(t_k;1) \qquad (4.18)$$

Correlation $corr\{pU^2\}$ of price and squares of trade volume depends on correlation $corr\{CU\}$ of the trade value *C* and trade volume *U,* on VWAP $p(t_k;1)$ and on trade volume volatility $\sigma^2(U)$. We repeat, that unnoticed zero correlations between price and trade volume as a consequence of VWAP introduced 30 years ago underlines importance of market-based approach to description of price random properties. Price should be studied as result of market trade (4.1) and that causes (4.4-4.10). Our definition (4.14-4.15) of price statistical moments $p(t_k;n)=E[p^n(t_i)]$ for all *n=1,2,3,…* causes zero correlations between *n-th* power of price $p^n(t_i)$ and trade volume $U^n(t_i)$. Relations (4.14; 4.15) define all price statistical moments $p(t_k;n)$ and hence completely determine price as random variable during $\Delta_k$ (4.3) as functions of statistical moments of the market trade value $C(t_k;n)$ (4.11) and volume $U(t_k;n)$ (4.12).

However, any econometric records of market trade during $\Delta_k$ (4.3) allow assess only finite number of market statistical moments $C(t_k;n)$ and $U(t_k;n)$. Thus, researchers should consider approximations of the market trade value and volume probabilities, determined by finite number of statistical moments. That causes assessment of finite number of the market price statistical moments those determine approximations of the price probability measure. Predictions of the price probability should follow forecasts of the market trade statistical moments. These ties between stochasticity of the market trade and randomness of the market price underline their economic nature and mutual dependence.

Predictions of the market trade *n-th* statistical moments for *n=2,3,...* uncover additional unnoticed variables and overlooked economic problems. Indeed, current economic theories consider variables composed by sums of the 1-st degree variables of economic agents. Macroeconomic investment, credits, assets are determined as sums of corresponding variables of agents of the entire economy. In turn, if one ignores impact of consumption and production, then agent's additive economic variables during interval $\Delta_k$ (4.3) are composed by sums of *1-st* degree market trade values $C_\Sigma(t_k;1)$ (4.11) and volumes $U_\Sigma(t_k;1)$ (4.12). Actually, sums of *n-th* power of market trade values $C_\Sigma(t_k;n)$ (4.11) and volumes $U_\Sigma(t_k;n)$ (4.12) during $\Delta_k$ (4.3) determine change of the macroeconomic variables of the *n-th* degree. These *n-th* degree macroeconomic variables establish direct ties between market stochasticity determined by market statistical moments $C(t_k;n)$ (4.11) and $U(t_k;n)$ (4.12) and unnoticed macroeconomic randomness. One should study evolution of *n-th* degree investment, credits, assets, because they record market stochasticity and project it into macroeconomic randomness. Evolution of the *n-th* degree macroeconomic variables depend on and impact on sums of *n-th* power of market trade values $C_\Sigma(t_k;n)$ (4.11) and volumes $U_\Sigma(t_k;n)$ (4.12). Their mutual interactions on one hand affect macroeconomic evolution and on other hand determine stochastic properties of the market trade statistical moments $C(t_k;n)$ and $U(t_k;n)$ (4.11; 4.12) and through them the market price statistical moments $p(t_k;n)$ (4.14; 4.15). Stochasticity of the market trade and price is intertwined into evolution of *n-th* degree economic variables. That is a really though puzzle for theoretical description.

Meanwhile, that "can of worms" formed by interrelations between macroeconomic variables, market trade and market price is only prelude, only preliminary introduction to the complexity of the comprehensive theory of economic reality. To imagine it to some extent one should remember that market trades to a large degree are conducted by collective agents expectations. Expectations, in their turn, are largely shaped and established under economic policy decisions and economic legislative environment, regulations and laws. In their turn, economic policy decisions are established by academic economists and politicians on base of their observations of macroeconomic variables their treatment of economic theory and their desires and goals to "improve" economic performance. Each novel policy decision disturbs and changes existing economic legislation and laws and cause perturbations of agents' expectations. In their turn disturbed expectations perturb market trade, market price and cause macroeconomic shocks. Actually, politicians supported by elite economists play one of the leading roles in regular perturbations of agents' expectations, market trade, price and, eventually, of macroeconomic performance.

## 5. Conclusion

It is clear that one paper cannot address and discuss all unnoticed issues and overlooked factors that may cause failures of the modern economic theories and policies. Many tough problems of economic modelling those arise within the proposed unified approach to description of economic variables, market transactions and expectations as functions of risk coordinates in the economic domain are left unnoticed. However, we assume that the above considerations of collective economic variables, collective transactions and expectations, mean risks of economic variables and transactions, collective velocities and flows of economic variables, transactions and expectations and other specific features of modelling their evolution and mutual interactions in the economic domain uncovers a great amount of unnoticed economic variables and overlooked economic processes those configure economic performance. The bounds of the *n*-dimensional economic domain result inherent cyclical motion of collective variables, transactions and expectations those are responsible for observed business cycles. Slow motion of collective economic variables and market transactions in the economic domain is complemented by propagation of numerous waves generated by small perturbations of different economic variables and market transactions. Ensemble of economic waves transfer perturbations over the economic domain and possible amplification of wave amplitudes can cause growth of economic and financial instabilities. Collective expectations those generally determine market trade and economic, financial, tax, trade and etc., laws and regulations those repeatedly disturbed by the next economic policy decisions complement that general view on vast variety of the theoretical economics puzzle.

The simplest way to solve the economic problems we mentioned above is to declare that all these complexities don't exist. Probably, that is the best choice for majority.

However, few readers may wonder in details of the results obtained and in the constructive proposal for the further development of the above problems and theoretical economics.

We believe that the general look on theoretical economics and numerous unnoticed and overlooked economic processes helps develop successive approximations of economic models piece by piece those complement each other. The comprehensive economic theory that can respond above queries is the goal of the far future. However, we propose that possible gains of econometric observations and theoretical description of the unnoted and overlooked economic factors and processes will reward any efforts.

We hope that our treatment of unnoticed and overlooked factors of theoretical economics and policy decisions preserves a wide field of studies for many decades for academic researchers, economic authorities and high-level politicians.